\documentclass{llncs}

\usepackage{amsfonts}
\usepackage{subfigure}
\usepackage{graphicx}
\usepackage{hyperref}
\let\epsilon=\varepsilon
\def\emph#1{\textit{\textbf{\boldmath #1}}}

\begin{document}

\title{Negative Instance for the \\ Edge Patrolling Beacon Problem}
\author{
Zachary Abel\inst{1} \and 
Hugo A. Akitaya\inst{2}\and
Erik D. Demaine\inst{1} \and
Martin L. Demaine\inst{1} \and
Adam Hesterberg\inst{1} \and
Matias Korman\inst{3}\and 
Jason S. Ku\inst{1} \and
Jayson Lynch\inst{1} 
}

\institute{Massachusetts Institute of Technology, Cambridge, USA,
\url{{zabel,edemaine,mdemaine,achester,jasonku,jaysonl}@mit.edu}
\and Carleton University, Ottawa, Canada, \url{hugoakitaya@gmail.com}
\and Tufts University, Boston, USA, \url{matias.korman@tufts.edu}
}

\maketitle

\begin{abstract}
  Can an infinite-strength magnetic beacon always ``catch'' an iron ball, when
  the beacon is a point required to be remain nonstrictly outside a polygon,
  and the ball is a point always moving instantaneously and maximally toward
  the beacon subject to staying nonstrictly within the same polygon?
  Kouhestani and Rappaport [JCDCG 2017] gave an algorithm
  for determining whether a ball-capturing beacon strategy exists,
  while conjecturing that such a strategy always exists.
  We disprove this conjecture by constructing orthogonal and general-position
  polygons in which the ball and the beacon can never be united.
\end{abstract}


\begin{figure}[htb]
  \centering
    \includegraphics[width=0.5\linewidth]{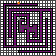}
    \caption{Orthogonal counterexample to Kouhestani and Rappaport's conjecture
      \cite{conj}.  The initial ball location is the grey dot;
      the initial beacon location is the orange dot;
      and the polygon is the purple region with black outline.}
    \label{fig_counter1}
\end{figure}

\section{Introduction}

\begin{figure}[htb]
  \centering
  \includegraphics[width=.8\linewidth]{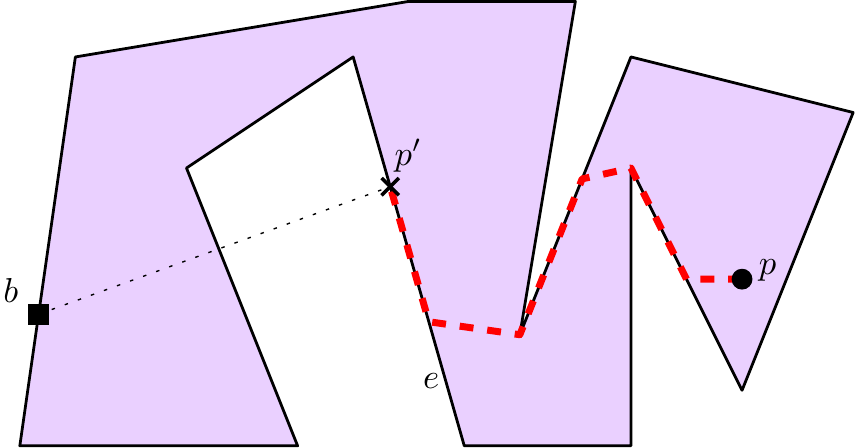}
  \caption{The beacon attraction model of \cite{Biro-FWCG,Biro-CCCG,jcdcg}.
    When the beacon $b$ is active, it attracts the ball $p$
    along the dashed red trajectory. The ball $p$ moves within the interior
    and boundary of $P$ while greedily reducing its Euclidean distance to $b$.
    Its movement stops when the ball reaches a point $p'$ (shown by a cross)
    on some edge $e$ of $P$
    for which no local movement can reduce its distance to~$b$.
    In particular, the segment $\overline{p'b}$ and the edge $e$ of $P$
    form a right angle.}
  \label{fig_example}
\end{figure}

Suppose you have a metallic iron \emph{ball}, modeled as a point
constrained to be within some compact polygonal region~$P$.
To manipulate the ball's location, you have one or more strongly magnetic
\emph{beacons}, also modeled as points.
When a beacon is \emph{active}, the ball instantaneously moves maximally
toward the beacon subject to staying within~$P$; refer to
\figurename~\ref{fig_example}.
More precisely, the ball moves toward the board in a straight line until it
either reaches the beacon or hits the polygon's boundary $\partial P$.
In the latter case, the ball continues gliding along $\partial P$
as long as the distance to the beacon decreases monotonically;
the ball may later leave the boundary $\partial P$ and resume moving along
a straight line to the beacon.
The ball stops moving either when reaching the beacon or
on a boundary edge perpendicular to the segment connecting the ball and beacon.
The general goal is to ``capture'' the ball by moving it to one of the beacons.

Biro et al.~\cite{Biro-FWCG,Biro-CCCG} introduced this model as a
generalization of the art gallery problem, inspired by geographical
greedy routing in sensor networks.
In their problem, beacons are stationary (cannot move)
and are only active when toggled into that state
(with at most one beacon active at any time).
The goal is to place the fewest possible beacons within $P$
to enable routing the ball from a given start point $s$ to a
given destination point $t$ (which will naturally have a beacon)
by activating beacons one at a time.
Similar to the classic art gallery problem,
they proved that this problem is NP-hard to solve exactly,
and proved that $\lfloor n/2 \rfloor - 1$ beacons are always sufficient
and sometimes necessary, where $n$ is the number of vertices of
a simple polygon~$P$.

Several papers have extended this initial work in various directions.
Biro's thesis \cite{Biro-thesis} details the results above
and improves the NP-hardness to APX-hardness.
By contrast, Biro et al.~\cite{Biro-WADS} give a PTAS when the beacons are
also allowed to be exterior to~$P$.
Shermer \cite{shermer2015combinatorial} analyzes the special case of
orthogonal polygons, where the worst-case bound improves to
$\lfloor {n-4 \over 3} \rfloor$.
Cleve and Mulzer \cite{Cleve-Mulzer-2018} analyzes the analogous problem
in 3D polyhedra.
Kostitsyna et al.~\cite{inverse-optimal} proves that $\Theta(n \log n)$ is
the optimal running time for computing the region of beacon positions
that can directly capture a given ball position in a given simple polygon.
Biro et al.~\cite{Biro-WADS} proves that this region has linear combinatorial
complexity but can have $\Omega(n)$ disconnected regions.

Kouhestani and Rappaport~\cite{jcdcg} introduced a different problem which is
the focus of our paper.
In their problem, there is a single beacon that is permanently active
and that can move along the boundary of~$P$.
The ball moves infinitely faster than the beacon, so it instantaneously
moves maximally toward the beacon as the beacon moves.
In addition to the polygon $P$, we are given the starting location of both
the ball $p$ (which could be either in the interior or on boundary of $P$)
and the beacon $b$ (which can only be on the boundary).
The goal is to give a beacon movement strategy, which moves the beacon
continuously along the boundary of~$P$ (but possibly changing directions),
so that the ball and beacon eventually coincide at the same point. 

Kouhestani and Rappaport~\cite{jcdcg} give an algorithm that, in $O(n^3\log n)$
time, determines whether such a beacon movement strategy exists.  Whenever
it exists, the algorithm also reports the shortest way for both objects to
meet (by measuring only the movement of the beacon, the ball, or both).
Although their algorithm determines the feasibility for any problem instance,
they could never design an instance for which their algorithm would return a
negative answer.  Thus they conjectured that all polygon instances have a
strategy for the beacon to reach the ball~\cite{conj}. 

\begin{figure}[htb]
  \centering
    \includegraphics[width=0.6\linewidth]{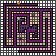}
    \caption{Orthogonal counterexample to Kouhestani and Rappaport's conjecture
      \cite{conj} from Figure~\ref{fig_counter1} with added details.
      The initial ball location is the grey dot;
      the initial beacon location is the orange dot;
      and the polygon is the (light and dark) purple region
      with black outline.
      The dark-purple region is called the \emph{core}, and the two
      orange/pink regions $W_1$ and $W_2$ are called \emph{pockets}.}
    \label{fig_counter}
\end{figure}

In this paper, we disprove this conjecture by giving a problem instance in which
the ball and the beacon can never be united. In fact, the same fact holds true
even if (1)~we allow the beacon to move freely in the exterior of $P$
(in addition to the boundary $\partial P$) and either
(2)~the polygon $P$ is orthogonal or
($2'$)~the polygon $P$ is in general position.
\figurename~\ref{fig_counter} shows the orthogonal counterexample;
the general-position example is a small perturbation thereof.
An interactive version of \figurename~\ref{fig_counter} and other examples
(implemented in a grid-based discrete model)
can be found at \url{http://erikdemaine.org/attractor/}. 

Our example is tight in the sense that no more freedom to
the movement of the beacon can be added: if we allow the beacon to move in the
interior of $P$, then the ball can be easily captured by walking along the
geodesic from the starting positions (a fact observed in~\cite{thesis}). 

\section{Grid-Orthogonal Counterexample}

\subsection{Construction}

Refer to Fig.~\ref{fig_counter}.
The key of the construction is the region shown in dark purple,
called the \emph{core},
that will always contain the ball. 
If we partition the core in the middle by a vertical line,
we obtain two rotationally symmetrical halves called \emph{hooks}.
The union of the shaded regions forms the orthogonal convex hull of~$P$.
The two orange/pink regions denoted $W_1$ and $W_2$ are called \emph{pockets}. 

The core has the following basic properties.
If we place the ball and beacon at the initial positions shown in \figurename~\ref{fig_counter}, and move the beacon clockwise along the boundary of the orthogonal convex hull of $P$, then the ball will always stay in the same hook.  
If we move the beacon counterclockwise along the boundary of the orthogonal convex hull, then the ball will instead alternate between both hooks (but will never leave the core). 

The main idea of the rest of the construction is to limit the possible paths between the beacon and the core so that, in order to reach the left hook (via $W_2$), the beacon must walk counterclockwise relative to the core which makes the ball move to the right hook, and vice versa for the other hook.
The shape of the pockets constrain the possible paths between the beacon and the core so that, even if we allow free movement in the exterior, the beacon cannot meet the ball.

\subsection{Correctness}

To be certain that there is no method for the two objects to meet (say, by going into a pocket then undoing the steps and going into another pocket), we verify all cases exhaustively as follows.

Define the \emph{grid space} to be the two-dimensional cell complex consisting of \emph{cells} --- corners, unit-length edges (excluding their endpoints), and unit-square faces (excluding their boundary) --- of the unit square grid,
and the \emph{beacon space} and \emph{ball space} to be the subspaces of the grid space that the beacon and ball can occupy, respectively.
(The ball space consists only of vertices and edges, i.e., zero- and one-dimensional cells, of the grid space.)
Because our example is orthogonal with boundaries on the unit square grid, as long as the beacon remains within one cell of its space, the ball will remain within one cell of its space (of lower or equal dimension), in particular because, when the ball rests on an edge of the grid, that edge is orthogonal to the line segment connecting the beacon and the ball.
Define the \emph{configuration space} to be the directed graph in which each node is a pair of a cell in the beacon space (\emph{beacon location}) and a cell in the ball space (\emph{ball location}), and we add a directed edge $(u,v)$ if the beacon location of $u$ is incident to the beacon location of $v$, and moving the beacon from the former to the latter causes the ball to move from the ball location of $u$ to the (not necessarily incident) ball location of~$v$.
We want to explore this graph and determine whether there is a node (1)~reachable from the starting position and (2)~where the ball and beacon are in a common cell of the grid space. 

\begin{figure}[tb]
  \centering
  \begin{tabular}{cc}
    \includegraphics[width=0.49\linewidth]{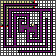} &
    \includegraphics[width=0.49\linewidth]{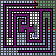} \\ 
    \includegraphics[width=0.49\linewidth]{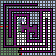} &
    \includegraphics[width=0.49\linewidth]{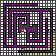}
  \end{tabular}
  \caption{Configuration space of pairs of positions reachable by the ball and beacon from the starting position in \figurename~\ref{fig_counter}. Reachable ball locations are the colored vertices and edges in the core, where colors correspond to labels. The figure is split into four images, each with disjoint regions for the beacon (outlined by solid edges and colored differently). Circle-labeled arrows denote transitions (``portals'') between regions from different images (along red edges, in the same direction as the arrow). The top and bottom rows are symmetric cases for the two hooks. These figures are computed automatically via a breadth-first search.}
  \label{fig:entering}
\end{figure}

Although there is a quadratic number of possible pairs of positions for the ball and beacon, only a very small amount is reachable from the starting position shown.
We \emph{label} each ball location with the two-dimensional cells incident to it.
From the starting position, the ball can be incident to only fourteen such two-dimensional cells, denoted by letters A to N in \figurename~\ref{fig:entering}.
The transitions between these cells, where the ball location has multiple labels, correspond to points where the ball might jump locations, and thus must be considered carefully.
For each reachable vertex of the configuration space, we give the same labels to the beacon location and color it accordingly in \figurename~\ref{fig:entering}.

The four images that form \figurename~\ref{fig:entering} encompass the whole range of reachable configurations.
The bottom-left image shows the possible locations of the ball in the left hook (labels A to F, each marked with a different color).
For each such location, we have highlighted the possible beacon locations (using the same color).
For example, if we know that the ball is in the left hook (labels A to F), and the beacon is in the topmost-rightmost corner, then it follows that the ball location must be solely labeled A.
The other three images depict the remaining pairs of reachable positions for the ball and beacon, and the transitions (``portals'') between these different parts of the configuration space.
In each image, white locations are impossible for the beacon to reach if we enforce the ball position to have one of the labels in that image.

Although the beacon location does not uniquely determine the ball location, there are at most three possible locations for the ball for any given beacon location. There are also few movements of the beacon that cause a change in the ball location (cell): the beacon must transition between a colored zone and the boundary between colored zones (drawn with solid lines).
For example, if we start in a configuration where the ball and beacon locations are solely labeled N, putting the beacon in the colored region in the top-left image of \figurename~\ref{fig:entering}, and then the beacon moves sufficiently to the right, then we reach a configuration where the ball and beacon locations are solely labeled M, represented in the top-right image of \figurename~\ref{fig:entering}.

Given the size of the instance, it is not difficult to verify that the images of \figurename~\ref{fig:entering} are correct. For example, if we start with the ball at a location solely labeled A and move the beacon around within the correspondingly colored beacon zone depicted in the bottom-left image of \figurename~\ref{fig:entering}, we can verify that the ball remains in a location solely labeled A. Indeed, the only ways of moving the ball to a differently labeled position are the four pictured transitions, and they all move the ball as denoted in the diagram. Because the maps from H to N are symmetric to A through G, we actually only need to check the seven zones corresponding to one row of \figurename~\ref{fig:entering}), which reduces to a small number of cases.

In addition to human verification, we have verified the correctness of our claim computationally using a breadth-first search.
The algorithm explores all states reachable from the initial state, and
verifies that no reachable state has the ball and beacon at the same location.
This shows that, as long as the ball and beacon start in any configuration denoted in \figurename~\ref{fig:entering}, the ball and beacon will never be able to unite, and furthermore that the ball cannot exit the two hooks.

\section{General Position Counterexample}

We claim we can modify our orthogonal counterexample into a polygon
in general position in which the beacon still cannot catch the ball.
Consider perturbing each polygon vertex to lie within an $\epsilon$-radius disk,
where $\epsilon$ is small enough to preserve the total order of
horizontal edge extensions and the total order of vertical edge extensions,
except when the original extensions are equal.
Specifically, if we set $\epsilon$ to one thousandth of a grid square, then
the angle of each edge changes by at most $\arctan \epsilon < 0.001$ radians,
so the total order of the 21 grid-square extensions are preserved
in both directions.

We claim that \figurename~\ref{fig:entering} still captures the beacon/ball
behavior of the perturbed example, with a perturbed notion of labeling.
Specifically, we define the boundaries between ball labels B--F according to
the extensions of the corresponding horizontal edges in the left hook,
with the exception of the transition between labels D and E, which we define
by the extension of the corresponding horizontal edge in the right hook
(as there is no corresponding horizontal edge in the left hook).
Symmetrically, we can define the boundaries between labels I--M.
For vertical transitions, we define the boundaries between labels A and B
and between labels F and G according to the extension of the corresponding
vertical edge in the left hook (the right edge of cell C),
and symmetrically between labels H and I and between labels M and N.
The boundaries between the beacon colored zones are defined by the extensions
of the same edges of the corresponding boundaries between ball labels.
Because the perturbation is small and our example is designed so that each
transition is affected by only one polygon edge (the one defined above),
the resulting configuration space is the same graph as in the orthogonal
polygon, with identical transition behavior.
Therefore the same arguments hold.

\section*{Acknowledgments}

We thank Dylan Hendrickson for helpful discussions,
and the anonymous referees for helpful comments.

\bibliography{beacon}
\bibliographystyle{plain}

\end{document}